\documentclass{JHEP3}

\usepackage{epsfig,amsmath,amssymb,multicol,bbm}

\newcommand{\beq}{\begin{equation}}
\newcommand{\eeq}{\end{equation}}
\newcommand{\ben}{\begin{eqnarray}}
\newcommand{\een}{\end{eqnarray}}
\newcommand{\bi}{\begin{itemize}}
\newcommand{\ei}{\end{itemize}}
\newcommand{\nn}{\nonumber}
\newcommand{\eg}{\textit{e.g.}}
\newcommand{\ie}{\textit{i.e.}}
\newcommand{\cf}{\textit{cf.}}

\newcommand{\mchi}{\mbox{$m_\chi$}}
\newcommand{\sigv}{\mbox{$\langle \sigma v \rangle$}}

\graphicspath{{./Figures/}}
\title{On the Sunyaev-Zel'dovich effect from dark matter annihilation or decay 
in galaxy clusters}

\author{Julien Lavalle\\ 
  Dipartimento di Fisica Teorica,
  Universit\`a di Torino \& INFN, via Giuria 1, 10125 Torino --- Italia\\
  Email: \email{lavalle@to.infn.it}}

\author{C\'eline B\oe hm\\
  LAPTH, UMR 5108, 9 chemin de Bellevue - BP 110,
  74941 Annecy-Le-Vieux --- France\\
  Email: \email{boehm@lapp.in2p3.fr}}

\author{Julien Barth\`es\\
  LAPTH, UMR 5108, 9 chemin de Bellevue - BP 110,
  74941 Annecy-Le-Vieux --- France\\
  Email: \email{barthes@lapp.in2p3.fr}}

\abstract{
  We revisit the prospects for detecting the Sunyaev Zel'dovich (SZ) effect 
  induced by dark matter (DM) annihilation or decay. We show that with 
  standard (or even extreme) assumptions for DM properties, the optical depth 
  associated with relativistic electrons injected from DM annihilation or 
  decay is much smaller than that associated with thermal electrons, when 
  averaged over the angular resolution of current and future experiments. For 
  example, we find: 
  $\tau_{\rm DM} \sim 10^{-9}-10^{-5}$ (depending on the assumptions) for 
  $\mchi = 1$ GeV and a density profile $\rho\propto r^{-1}$ for a template 
  cluster located at 50 Mpc and observed within an angular resolution of 
  $10''$, compared to $\tau_{\rm th}\sim 10^{-3}-10^{-2}$. This, together with 
  a full spectral analysis, enables us to demonstrate that, for a template 
  cluster with generic properties, the SZ effect due to DM annihilation or 
  decay is far below the sensitivity of the Planck satellite. This is at 
  variance with previous claims regarding heavier annihilating DM particles. 
  Should DM be made of lighter particles, the current constraints from 511 keV 
  observations on the annihilation cross section or decay rate still prevent a 
  detectable SZ effect. Finally, we show that spatial diffusion sets a core of 
  a few kpc in the electron distribution, even for very cuspy DM profiles, such
  that improving the angular resolution of the instrument, \eg~with ALMA, does 
  not necessarily improve the detection potential. We provide useful 
  analytical formul\ae~parameterized in terms of the DM mass, decay rate or 
  annihilation cross section and DM halo features, that allow quick estimates 
  of the SZ effect induced by any given candidate and any DM halo profile.}


\preprint{DFTT 48/2009}

\begin{document}

\section{Introduction}
\label{sec:intro}
The enigma of the origin of dark matter (DM) is a longstanding issue. Many 
experimental strategies in particle physics, astrophysics and cosmology 
have been developed to confirm/infirm the existing scenarios. One appealing 
possibility, strongly motivated by independent issues in particle physics 
beyond the standard model (BSM), is that DM is made of weakly interacting 
particles (WIMPs). The most popular models, like the supersymmetric (SUSY) 
neutralino (see e.g.~\cite{1996PhR...267..195J,2007arXiv0704.2276M} for 
reviews), have the interesting property of self-annihilation to ordinary 
matter, which provides windows for the detection of the annihilation products. 
There is actually an impressive wealth of WIMP models, from the GeV-TeV 
mass-scale as in the SUSY paradigm down to the sub-GeV mass-scale, the latter 
class allowing, in particular, to solve the so-called \emph{cusp} and 
\emph{subhalo problems} (\eg~\cite{2004NuPhB.683..219B}). Since the DM 
annihilation or decay rate increases with the DM number density, the best 
targets are expected to be the centers of (sub)galaxies or galaxy clusters. 
This quest for annihilation or decay signals makes use of different messengers
(\eg\ $\gamma$-rays, antimatter cosmic rays and neutrinos): it is referred to 
as \emph{indirect detection} of DM (for a pedagogical review, 
see~\cite{Salati:2007zz}). There are of course other ways 
to learn about the microscopic properties of DM, for instance by detecting 
direct interactions of DM in ground based experiments~\cite{Salati:2007zz}, 
or by producing DM directly in particle colliders. All these direct and 
indirect detection methods have been studied at length in the literature.

Yet, a few years ago, it was pointed out in a series of papers
\cite{2001ApJ...562...24C,2004A&A...422L..23C,2006A&A...455...21C} a new 
complementary method to look for DM, namely the Sunyaev-Zel'dovich (SZ) 
effect~\cite{1972CoASP...4..173S,1980ARA&A..18..537S}. Indeed, DM annihilation 
or decay in galaxy clusters may inject relativistic electrons and 
positrons\footnote{\label{foot:1} Hereafter, we will use \emph{electron} for 
electron or positron, indifferently.} that should experience inverse Compton 
scattering with the CMB photons and, therefore, if numerous enough, generate a 
deviation to the blackbody spectrum. The question that we address in this paper 
is whether or not this deviation is observable in light of the contradictory 
results found by 
\cite{2001ApJ...562...24C,2004A&A...422L..23C,2006A&A...455...21C} on the 
one hand, and by~\cite{2009JCAP...10..013Y} on the other hand. In particular, 
we investigate whether the Planck satellite~\cite{2006astro.ph..4069T}, which 
was recently launched, and the ALMA 
facilities~\cite{alma_paper,2009arXiv0904.3739W} will have the sensitivity to 
detect the SZ signal induced by DM annihilation or decay and could constrain 
the DM properties.

Our approach aims at providing more analytical insights 
than~\cite{2009JCAP...10..013Y} on the crucial impact of the experimental 
angular resolution on the DM-induced SZ predictions, on the spectral 
analysis itself and on spatial diffusion effects. We also consider the SZ 
effect due to the thermal electrons lying in clusters and observed in X-ray 
measurements, which has been studied for a long time 
(see~\cite{1999PhR...310...97B} for a review) and 
which constitutes an important background to such an exotic signature.
To proceed, we will consider a template galaxy cluster located at 50 Mpc, with 
properties representative of all nearby clusters. Our discussion, that we will 
keep as generic as possible, is 
organized as follows. In Sec.~\ref{sec:principle}, we will recall the principle 
of the SZ calculation. In Sec.~\ref{sec:dm_el} we will derive the optical 
depths for both DM-induced and thermal electrons in an exact numerical 
manner and, likewise, provide analytical formulae to allow quick calculation
of the SZ effect for any cluster and DM properties. In Sec.~\ref{sec:res}, we 
will then use these results together with a full spectral analysis to compute 
the signal due to the DM-induced electrons at the transition frequency where 
the thermal component is negligible, and show that the DM contribution to the 
SZ will be quite hardly observable with current and even future radio 
experiments. In Sec.~\ref{sec:diff}, we will close the case by considering 
spatial diffusion effects, showing analytically that they induce a core in
the electron distribution whatever the cuspiness of the DM profile. We will 
conclude in Sec.~\ref{sec:concl}.

\section{Principle}
\label{sec:principle}

There are basically two main approaches to calculate the thermal or 
relativistic SZ effect, the radiative transfer method proposed in 
\eg~\cite{1979ApJ...232..348W} and early formalized 
in~\cite{1950ratr.book.....C}, and the covariant Boltzmann 
equation formalism~\cite{1998ApJ...499....1C,1998ApJ...502....7I,1998ApJ...508....1S,1997astro.ph..9065S,2001ApJ...554...74D}. Recently, however, these 
approaches have been demonstrated to be 
equivalent~\cite{2009PhRvD..79h3505B,2009PhRvD..79h3005N}, which settles a 
self-consistent and unique framework for the SZ calculations. In the single 
scattering approximation, the distortion of the CMB intensity $I$ at 
a photon energy $E_k$ is given by the following equation:
\ben
\Delta I_\gamma (E_k) &=& 
\tau \int_0^{t_{\rm max}} dt\; \left( P_1(t) - \delta(t-1) \right)
I_\gamma^0 (E_k/t)
\label{eq:deltaI}
\een
where $I_\gamma^0 $ is the undistorted blackbody intensity, $P_1$ is the 
frequency redistribution function for a single electron-photon interaction and 
$t$ is the ratio of scattered to unscattered photon energy. The integrand 
encodes the spectral features: the first term describes the spectral part 
that is shifted from energy $E_k/t$ to $E_k$, while the second term describes 
the part that is removed from energy $E_k$ to others. $\tau$, which weights 
the amplitude of the distortion, is the optical depth which is related to the 
electron density in the cluster through the line-of-sight integral
\ben
\tau \equiv \sigma_{\rm T} \int dl \, n_e(\vec{x})\;.
\label{eq:def_tau}
\een
Here $\sigma_{\rm T}$ is the non-relativistic Thomson cross section that fully 
characterizes interactions of CMB photons with electrons of Lorentz factor 
$\gamma_e \lesssim 10^8$. As far as $\tau \ll 1$, the single scattering 
approximation is fully justified~\cite{2009PhRvD..79h3005N}.

In order to estimate the potential of a radio experiments to detect SZ signals, 
however, one needs to average the optical depth within the angular resolution 
of the apparatus,
\ben
\langle \tau \rangle_{\mu_{\rm res}} &=& 
\frac{\sigma_{\rm T}}{\Delta\Omega(\mu_{\rm res})}
\int_{\Delta\Omega(\mu_{\rm res})} d\Omega \int dl \, n_e(\vec{x})\;,
\label{eq:def_tau_av}
\een
where $\mu_{\rm res}\equiv \cos(\theta_{\rm res})$ characterizes the angular 
resolution $\theta_{\rm res}$.

Disregarding for the moment the spectral aspects, the observation of a
relativistic SZ signal on top of the thermal contribution would roughly imply 
that the optical depth of the additional electron component is sizable 
compared to the one associated with the thermal electrons. A mere comparison 
of the electron densities in the cluster center, as done 
in~\cite{2009JCAP...10..013Y}, is indicative at $0^{\rm th}$ order.
Nevertheless, because of the angular averaging, this could be misleading.
Indeed, the SZ signal is in any case smeared down by angular resolution effects.

In Sec.~\ref{sec:dm_el}, we will study the effects of this averaging
procedure, which is actually very common in studies on the indirect detection 
of DM with $\gamma$-rays (\eg~\cite{1998APh.....9..137B}). 
We will consider angular resolutions between $\sim 0.1''$ and $1'$, noticing 
that Planck~\cite{2006astro.ph..4069T} and 
ALMA~\cite{alma_paper,2009arXiv0904.3739W} could reach resolutions of 
$\sim5'$ and $\sim 1''$, respectively.

Throughout the paper, moreover, we will consider the spatial distribution of 
the thermal electrons to be a spherical cored isothermal:
\ben
n_{e,{\rm th}}(r) = 
\frac{n_{e,{\rm th}}^0}{ 1+\left(\frac{r}{r_c}\right)^2}\;,
\label{eq:th_el}
\een
where we will fix $n_{e,{\rm th}}^0$ to $0.01\,{\rm cm^{-3}}$, and 
$r_c$ to 400 kpc. These are typical values for nearby clusters that will 
partly characterize our \emph{template} cluster.

For completeness, it is worth recalling that the SZ effect is featured by the 
shift of the low frequency part of the CMB spectrum towards higher 
frequencies, but much more marginally by that of the high frequency part 
towards lower frequencies. This asymmetry drives $\Delta I$ to be negative at 
low frequencies and positive at high frequencies. The transition frequency, at 
which $\Delta I = 0$, is intimately related to the energy spectrum of the 
involved electron gas. Because the thermal and DM-induced components are found 
in different energy ranges, the associated transition frequencies are 
different. Therefore, not only do we need to compare the optical depths of 
both contributions, that weight the amplitudes of the corresponding 
distortions, but also to study the spectral features of $\Delta I$.
For the spectral study, we will mainly focus on $\Delta I_\chi$ at the 
transition frequency of the thermal component ($\Delta I_{\rm th} = 0$) in 
Sec.~\ref{sec:res}.

\section{Injected electron density from DM annihilation or decay and 
optical depths}
\label{sec:dm_el}

The injection rate of relativistic electrons (and positrons) from DM 
annihilation or decay characterizes a nominal source term
\ben
{\cal Q}_{n,\gamma}(E,\vec{x}) = N_0 \,\alpha_n \, 
\left( \frac{\rho_\gamma(\vec{x})}{\mchi} \right)^n \, {\cal F}(E)\;,
\label{eq:source_dm}
\een
where $N_0$ is the total number of electrons and positrons injected per 
annihilation or decay in the relevant energy range, ${\cal F}(E)$ is the 
energy spectrum normalized to unity, and $\mchi$ is the DM particle mass. The 
index $n$ is 1 or 2 for DM decay or annihilation, respectively. Therefore, 
$\alpha_1 \equiv \Gamma_\chi$ is a decay rate, while 
$\alpha_2 \equiv {\delta } \sigv/2$ ($\delta = 1$ or $1/2$ for Majorana 
or Dirac fermions, $1$ for bosons) is an annihilation rate. The 
DM mass density profile $\rho_\gamma$ is indexed by its inner logarithmic 
slope $\gamma$, and is usually written like a spherical symmetric component 
with a scale radius $r_s$ and a scale density 
$\rho_s$~\cite{1996MNRAS.278..488Z}
\ben
\rho_\gamma(r) = \frac{ \rho_s \, ( r_s/r )^\gamma}
{\left( 1 + (r/r_s)^\alpha \right)^{(\beta-\gamma)/\alpha}} = 
\rho_s \,f_\gamma(r) \;.
\label{eq:dm_prof}
\een
The so-called Navarro-Frenk-White (NFW) profile~\cite{1997ApJ...490..493N} 
corresponds to $(\alpha,\beta,\gamma)=(1,3,1)$, while the Moore 
profile~\cite{1998ApJ...499L...5M} is even more cuspy with 
$(1.5,3,1.5)$. Notice that a cored isothermal profile with 
$(2,2,0)$ is equivalent to the thermal electron distribution of 
Eq.~(\ref{eq:th_el}), with $r_s=r_c$; this may be kept in mind with 
further benefit. We will study all of the three aforementioned profiles in the 
following, for both annihilating and decaying DM models. The typical values of 
the scale parameters found for galaxy clusters with masses $\sim 10^{15}M_\odot$
in cosmological N-body simulations are $r_s \sim 400$ kpc and 
$\rho_s \sim 0.05\,{\rm GeV.cm^{-3}}$ (\eg~\cite{2001MNRAS.321..559B}). We will 
adopt these parameters for the \emph{template} cluster that we will use for 
our calculations throughout this paper. We summarize them in Tab.~\ref{tab:cl}.

\TABLE{%\begin{table}
\centering
\begin{tabular}{ccccc}
\hline
distance & $R_{\rm vir}$ & $r_s$ and $r_c$& $\rho_s$ & thermal $e^-$ density\\
(Mpc) & (kpc) & (kpc) & (GeV/cm$^3$) & (${\rm cm^{-3}}$)\\
50 & 2000 & 400 & 0.05 & 0.01\\
\hline
\end{tabular}
\caption{Properties of the \emph{template} cluster used in this paper.}
\label{tab:cl}
}%\end{table}

After their injection in the intracluster medium, relativistic electrons and 
positrons diffuse in space and momentum. The main processes that come into 
play are the energy losses, with a typical timescale $\sim 300$ Myrs, 
and the spatial diffusion due to the scattering on the magnetic inhomogeneities.
It was shown in previous analyses that since the relevant spatial diffusion 
scale is $\sim$ kpc, which is just a bit larger than typical resolution 
scales, one can neglect spatial diffusion at first 
order~\cite{2006A&A...455...21C}. We will stick to this approximation, for 
which, in steady state, the diffusion equation reduces to
\ben
\frac{\partial}{\partial E} \left\{ b(E) 
\frac{dn_{e,\chi}^{n,\gamma}(E,\vec{x})}{dE} 
\right\} = {\cal Q}_{n,\gamma}(E,\vec{x}) \;,
\label{eq:diff_eq}
\een
where $b(E) \equiv -dE/dt$ is the energy loss rate. This equation is easily 
solved:
\ben
\frac{dn_{e,\chi}^{n,\gamma}(E,\vec{x})}{dE} &=& \frac{1}{b(E)}
\int_E^{\frac{n\,m_\chi}{2}} dE_s \,{\cal Q}_{n,\gamma}(E_s,\vec{x}) \\
&=& \frac{N_0 \,\alpha_n }{b(E)} 
\left( \frac{\rho_\gamma(\vec{x})}{\mchi} \right)^n 
\int_E^{\frac{n\,m_\chi}{2}}dE_s{\cal F}(E_s)\;.\nn
\label{eq:dnde_dm}
\een
Now, to compute the optical depth, as defined in Eq.~(\ref{eq:def_tau}), we 
need to integrate this differential electron density over energy. Using 
the previous equation, this gives:
\ben
\label{eq:nel_noav0}
n_{e,\chi}^{n,\gamma}(\vec{x}) = N_0 \,\alpha_n  
\left( \frac{\rho_\gamma(\vec{x})}{\mchi} \right)^n \,
\int_{E_{\rm min}}^{\frac{n\,m_\chi}{2}} \frac{dE}{b(E)} 
\int_E^{\frac{n\,m_\chi}{2}}dE_s{\cal F}(E_s)\;.
\een
We can further express the energy loss rate as 
$b(E) = (E_0/\tau_{\rm loss})/g(E)$, where $E_0 = 1$ GeV, $\tau_{\rm loss}$ 
is the typical energy loss timescale, and $g(E)$ is a dimensionless 
function that encodes the energy dependence of the energy loss rate. The 
previous equation is then more conveniently rewritten as
\ben
n_{e,\chi}^{n,\gamma}(\vec{x}) &=& N_0 \,\alpha_n  
\left( \frac{\rho_\gamma(\vec{x})}{\mchi} \right)^n \, \tau_{\rm loss} \, 
\bar{\cal F}\;,
\label{eq:nel_noav1}
\een
where we have defined
\ben
\bar{\cal F} \equiv \int_{E_{\rm min}}^{\frac{n\,m_\chi}{2}} \frac{dE}{E_0} g(E)\,
\int_E^{\frac{n\,m_\chi}{2}}dE_s{\cal F}(E_s)\;.
\label{eq:def_fbar}
\een
Note that $\bar{\cal F} \lesssim \mchi/E_0$ for Coulomb losses ($g(E) \propto 
{\rm cst}$), while $\bar{\cal F} \lesssim E_0/E_{\rm min}$ for inverse Compton 
losses ($g(E)\propto (E/E_0)^{-2}$), which is in any case $\lesssim 10^3$.

Now, to go further in our calculation of the optical depth, we need to 
compute the average density of electrons within the angular resolution of the 
telescope.

\subsection{Line-of-sight averaged optical depth and electron density}
\label{subsec:los_ave}

As stated in Sec.~\ref{sec:principle} through Eq.~(\ref{eq:def_tau_av}), 
the optical depth must be averaged within the experimental angular 
resolution. The full line-of-sight integral is not analytical in most of cases.
We can reexpress Eq.~(\ref{eq:def_tau_av}) in such a way that an averaged 
electron density explicitly appears:
\ben
\langle\tau_{e,\chi}^{\gamma,n}\rangle = 2\, r_s \, \sigma_{\rm T} \,
\widetilde{\langle n_{e,\chi}^{n,\gamma} \rangle}_{\rm res}\;.
\label{eq:tau_dm_av}
\een
Factorizing the scale radius $r_s$ out just translates the fact that the most 
important contribution to the SZ is expected to come from within a radius of 
$r_s$ from the cluster center (therefore within a distance of $2\,r_s$ along
the line of sight). This is essentially true for annihilating DM, but 
the equation is anyway made exact for all cases by defining the average 
electron density as
\ben
\widetilde{\langle n_{e,\chi}^{n,\gamma} \rangle}_{\rm res} &=& 
N_0 \,\alpha_n  
\left( \frac{\rho_s}{\mchi} \right)^n \, \tau_{\rm loss} \, 
\bar{\cal F} \, {\cal J}^{n,\gamma}\;.
\label{eq:av_nel_exact}
\een
This ensures Eq.~(\ref{eq:tau_dm_av}) to be an exact formulation of the 
angular averaging of the optical depth, provided a proper definition of the 
dimensionless parameter ${\cal J}^{n,\gamma}$, which carries the angular average:
\ben
\label{eq:Jlos}
{\cal J}^{n,\gamma} &\equiv& \frac{1}{2\,r_s\,(1-\mu_{\rm res})} 
\int_{\mu_{\rm res}}^{1}d\mu \int dl \,\left(f_\gamma (r(l,\mu))\right)^n \\
&=&\frac{1}{2\,r_s\,(1-\mu_{\rm res})} 
\int_{\mu_{\rm res}}^{1}d\mu \,
2\int_{0}^{\sqrt{R_{\rm vir}^2 - b^2}}ds\,  \left(f_\gamma(r(s,b)) \right)^{n}\nn\\
&=& \frac{(1+\mu_{\rm res})}{b_{\rm res}^2} \int_{0}^{b_{\rm res}} 
\frac{db\,b}{\sqrt{1-\frac{b^2}{D^2}}} \,
\int_{0}^{\sqrt{R_{\rm vir}^2 - b^2}} 
 \frac{ds}{r_s}\, \left(f_\gamma( \sqrt{ s^2+b^2 } ) \right)^{n} \;.\nn
\een
$b = D\sin(\theta_{\rm res})$ is the impact parameter, 
$\mu_{\rm res} = \cos(\theta_{\rm res})$, $f_\gamma(r) \equiv \rho_\gamma(r)/\rho_s$, $s\equiv\sqrt{r^2 -b^2}$ is the reduced line-of-sight variable and
$R_{\rm vir}$ is the virial radius of the cluster. We have expressed
the angular integral in terms of an integral over the impact parameter, which 
allows a better control of numerical aspects. It is usually convenient 
to integrate over $b$ and $s$ with logarithmic steps, and to define a numerical 
cut $r_{\rm cut}$ for $f_\gamma$ when $b\rightarrow 0$. Although formally set 
in a pure DM halo by equating the gravitational infall timescale with the 
annihilation timescale~\cite{1992PhLB..294..221B}, $r_{\rm cut}$ can be chosen 
as to ensure a good compromise between numerical accuracy, convergence and 
computing time, but should anyway obey $r_{\rm cut}\ll b_{\rm res}$. We 
remind, nevertheless, that small core radii could instead arise from other
non-trivial dynamical effects involving baryons, and are actually observed at 
the galactic scale~\cite{2004MNRAS.351..903G}.

With the previous definitions, simple estimates of the optical depths 
associated with DM decay and annihilation read, respectively:
\ben
\label{eq:dm_numbers}
\langle \tau_{e,\chi}^{1,\gamma} \rangle_{\rm res}^{\rm dec} &=& 
\tau_{0,\chi}^{\rm dec} \,  
\frac{N_0}{10} \, 
\frac{\bar{\cal F}}{10^3} \, 
\frac{\Gamma_\chi }{10^{-26}{\rm s^{-1}}} \, 
\frac{\tau_{\rm loss}}{10^{17}{\rm s}} \,
\left[ \frac{\rho_s /(0.05\,{\rm GeV/cm^3})}
     {\mchi /{\rm GeV}}\right] \,
\left[ \frac{2\,r_s\,{\cal J}^{1,\gamma}}{10^3\,{\rm kpc}} \right] \\
\langle \tau_{e,\chi}^{2,\gamma} \rangle_{\rm res}^{\rm ann} &=&
\tau_{0,\chi}^{\rm ann}\, 
\frac{N_0}{10} \, 
\frac{\bar{\cal F}}{10^3} \, 
\frac{\sigv }{3\cdot 10^{-26}{\rm cm^3 s^{-1}}} \, 
\frac{\tau_{\rm loss}}{10^{17}{\rm s}} \,
\left[ \frac{\rho_s /(0.05\,{\rm GeV/cm^3})}
     {\mchi /{\rm GeV}}\right]^2 \,
\left[ \frac{2\,r_s\,{\cal J}^{2,\gamma}}{10^5\,{\rm kpc}} \right]\;,\nn
\een
where the typical optical depths are given by
\ben
\tau_{0,\chi}^{\rm dec} = 1.01\times 10^{-6} \;\;\;{\rm and}\;\;\;
\tau_{0,\chi}^{\rm ann} = 7.68\times 10^{-6}\;.
\label{eq:typ_tau_dm}
\een
It is important to note that the values that we have taken for $N_0$ and 
$\bar{\cal F}$ typify mostly DM particles with masses above $\sim 100$ MeV, 
since for lighter particles, the annihilation or decay will be mostly into 
light fermion pairs. For direct annihilation or decay in $e^+e^-$, we would 
have $N_0 =2$ and $\bar{\cal F} \approx 1$ (see Sec.~\ref{sec:res}), which 
would reduce the previous estimates of the optical depths by a factor of 
$5\times 10^3$.

In Fig.~\ref{fig:res}, we plot the full results. We have sketched the 
behaviors of the exact computation of 
$\widetilde{\cal J}^{n,\gamma}\equiv 2 \, r_s {\cal J}^{n,\gamma}$ on the one 
hand, and of the corresponding analytical spherical / line-of-sight 
approximations derived in Sec.~\ref{subsec:los_approx} on the other hand, as 
functions of the resolution angle (or the corresponding impact parameter, 
which allows to deal with physical distances --- see the top horizontal axis). 
These results are translated in terms of optical depth by means of
the DM properties used in Eq.~(\ref{eq:dm_numbers}).
Note that larger values of $R_{\rm vir}$ would not affect the results, since 
the denser part of the DM distribution is within $r_s$. Different values 
of the cluster distance $D$ would affect the scaling of horizontal axis, 
the closer the cluster the better the spatial resolution.

The values of the DM-induced electron optical depths derived in 
Eq.~(\ref{eq:dm_numbers}) have to be compared with what is expected for the 
thermal electrons. Because the latter obey an isothermal cored profile, as 
made explicit in Eq.~(\ref{eq:th_el}), the 
function ${\cal J}^{1,0}$ can be used to infer their optical depth (we take 
advantage of that we have used $r_c = r_s$). We find:
\ben
\langle \tau \rangle_{\rm res}^{\rm th} = \tau_{0,{\rm th}}\, 
\left[ \frac{n_{e,{\rm th}}^0}{0.01\,{\rm cm^{-3}}}\right]
\left[ \frac{2\,r_c\,{\cal J}^{1,0}}{10^3\,{\rm kpc}} \right]\;,
\label{eq:th_numbers}
\een
where the typical optical depth is found to be
\ben
\tau_{0,{\rm th}} = 2.05\cdot 10^{-2}
\een
with our template parameters. Such a value is very close to what 
is actually found in dedicated studies, where 
$\tau\sim 10^{-3}-10^{-2}$~\cite{1999PhR...310...97B}. Moreover, this value 
remains a constant function of the angular resolution up to an angular size 
that corresponds, roughly, to the core radius $r_c \sim 400$ kpc.

If at first glance, still disregarding the spectral aspects, we want to compare 
the thermal and DM optical depths, then we can use Fig.~\ref{fig:res} (the 
left vertical axis gives $2 r_s {\cal J}^{n,\gamma}$ and the right vertical 
axis provides the translation in terms of optical depth for annihilating DM 
with the generic parameters used in Eq.~\ref{eq:dm_numbers}). From this 
figure, we can already emphasize that unless 
$\sigv \gg 3\cdot 10^{-26}{\rm cm^3/s}$, $\Gamma_\chi\gg 10^{-26}{\rm s^{-1}}$
or $\mchi\ll 1$ GeV, $\gamma > 1$ or $\rho_s \gg 0.05 \, {\rm GeV/cm^3}$, DM 
decay or annihilation cannot supply a sufficient amount of electrons to 
compete with the thermal optical depth. This even holds for rather light DM 
particles in the sub-GeV mass-scale given conventional values for the other 
parameters. Such a statement is in agreement with~\cite{2009JCAP...10..013Y}, 
though based on more striking analytical calculations, but in clear 
disagreement 
with~\cite{2001ApJ...562...24C,2004A&A...422L..23C,2006A&A...455...21C}.
The detection potential of future experiments with angular resolutions
$\sim 1''$ seems therefore very weak except for light DM with $\mchi < 1 $ GeV 
either associated with large annihilation or decay rates, or distributed with 
very cuspy profiles ($\gamma \gtrsim 1.5$). This latter feature is, however, 
strongly disfavored by most of the recent cosmological N-body simulations 
(e.g.~\cite{2008arXiv0810.1522N}).

Before tackling the full spectral analysis, we derive hereafter some 
approximated analytical expressions of the line-of-sight integral 
${\cal J}^{n,\gamma}$. This set of \emph{fast} formulae is aimed at providing 
some quick estimates of the optical depth for any annihilating or decaying DM 
model, and for any density profile discussed below Eq.~(\ref{eq:dm_prof}).

\subsection{Analytical approximations}
\label{subsec:los_approx}

The relevant part of the DM distribution is that within the angular 
resolution of the telescope. The current and foreseen angular resolutions are 
not better than $\theta_{\rm res}\sim 1''$. Since nearby clusters lie at 
distances around 50 Mpc, the spatial resolution can hardly be better than
$b_{\rm res} \sim D\,\theta_{\rm res} \sim 0.1$ kpc ($b$ is defined as the 
impact parameter). Such values are much 
smaller than the typical scale radii $r_s$ found for cluster density 
profiles, so we can approximate the density profile of Eq.~(\ref{eq:dm_prof}) 
as
\ben
\rho_\gamma (r) \approx  \rho_s \left( \frac{ r_s}{r }\right)^\gamma\;\;
\Leftrightarrow\;\; f_\gamma(r) \approx \left( \frac{ r_s}{r }\right)^\gamma\;.
\label{eq:approx}
\een
Beside the fact that the impact parameter is, in most cases, smaller than 
the scale radius $r_s$, this approximation is further justified from that the 
dominant part of the line-of-sight DM contribution arises within 
a sphere of radius $r_s$ by the cluster center (the argument is 
weaker for decaying DM). We will therefore disregard the whole spatial 
extent of the cluster characterized by $R_{\rm vir}$, and focus only on the 
region within $r_s$. We further assume $D\gg r_s \gg b_{\rm res}\gg r_{\rm cut}$.
This latter cut-off radius $r_{\rm cut}$ has already been discussed below 
Eq.~(\ref{eq:Jlos}). It has no impact except for diverging injection rates, 
scaling like $r^{-\alpha}$ with $\alpha\geq 3$ (corresponding for instance to 
annihilating DM distributed with a Moore profile); we will use 
$r_{\rm cut} = 10^{-3}$ pc in the following. We will develop two types of 
approximations, one based on a spherical average, the other based on a more
accurate line-of-sight treatment.

\subsubsection{The spherical approximation}
\label{subsubsec:sph_approx}

If the impact parameter is sufficiently small, the average value of the 
electron density determined in Eq.~(\ref{eq:av_nel_exact}) can be approximated 
by a spherical integral of $n_{e,\chi}$ around $b_{\rm res}$, weighted by the 
conic volume $\sim \pi \,b_{\rm res}^2\,R_{\rm vir}/3$ of the cluster roughly 
carried by the telescope resolution. In this case, we can derive a spherical 
approximation of the dimensionless function ${\cal J}^{n,\gamma}$ as follows
\ben
{\cal J}^{n,\gamma} \approx  {\cal J}^{n\gamma}_{\rm sph}
\approx 
\frac{12}{b_{\rm res}^2 R_{\rm vir}} \frac{R_{n,\beta,\gamma} }{r_s} r_s^{n\gamma}
\begin{cases} 
\frac{r^{3-n\gamma}}{3-n\gamma} |^{b_{\rm res}}_{r_{\rm cut}} \;,\; 
  n\gamma\neq 3\\
\ln\left( \frac{b_{\rm res}}{r_{\rm cut}}\right)\;\;{\rm otherwise}
\end{cases} \;.
\label{eq:sph_approx}
\een
A given product of $n\gamma$ corresponds to one or two DM configurations, 
as recalled in Tab.~\ref{tab:ngamma}.
The scaling with $b_{\rm res}$ is found to match exactly the actual behavior 
of ${\cal J}^{n,\gamma}$, but we need to define and tune a new scale, 
$R_{n,\beta,\gamma}$, to get closer to the actual value of the optical depth 
($\beta = 3$ for the NFW and Moore profiles, and $\beta = 2$ for cored 
isothermal profiles). $R_{n,\beta,\gamma}$ should be of the order $r_s$ for 
consistency reasons (see Eqs.~\ref{eq:tau_dm_av} and~\ref{eq:av_nel_exact}). 
We actually find that
\ben
R_{n,\beta,\gamma} = r^0_{n,\gamma}
\left(\frac{b_{\rm res}}{r_{n,\gamma}} \right)^{\beta-3}
\een
with
\ben
r^0_{1,\gamma} = r_s \;\; ; \;\; r_{1,\gamma} = \frac{r_s}{7} 
\;\;\;{\rm and}\;\;\; r^0_{2,\gamma} =  r_{2,\gamma} = r_s\;,
\een
provides a reasonable approximation of the accurate numerical results, by less 
than one order of magnitude for cluster parameters not too far from those of 
our template model. This is therefore consistent with the values expected for 
$R_{n,\beta,\gamma} $. The comparison with the accurate result is shown in 
Fig.~\ref{fig:res}. Although this approximation allows to derive orders of 
magnitude quite quickly, we recall that it is very simplistic, so one 
should use it with care. We recommend not to use it for very detailed analyses,
but instead to make the full computation of Eq.~(\ref{eq:tau_dm_av}). The 
line-of-sight approximation that we propose in the next paragraph turns out 
to be much more accurate; still, the spherical case is rather intuitive and 
provides complementary insights.

\FIGURE[t]{%\begin{figure}[t]
%\begin{center}
\includegraphics[width=\columnwidth, clip]{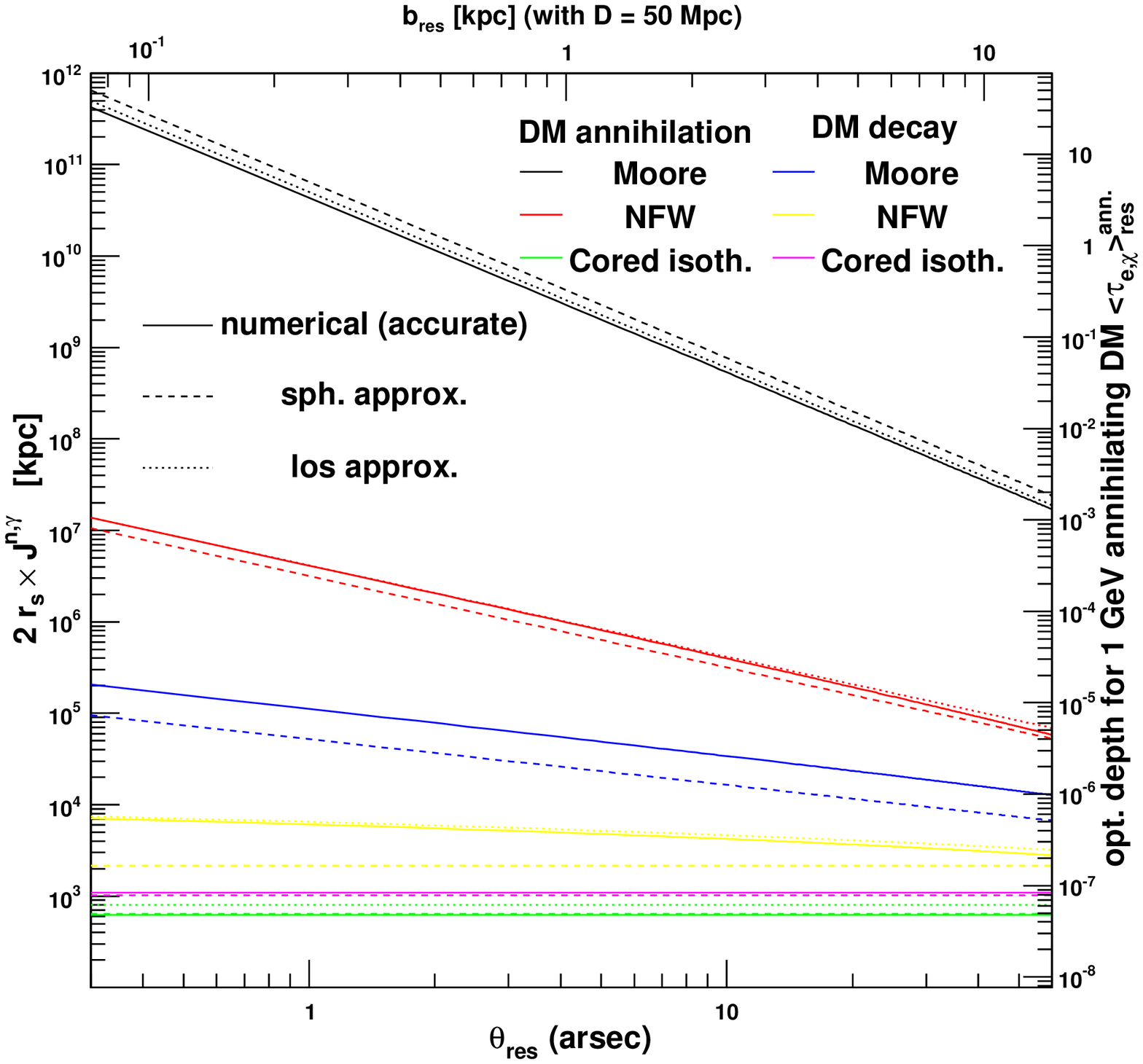}
\caption{Line-of-sight integral for the accurate numerical computation, 
the spherical approximation and the so-called line-of-sight approximation, 
as a function of the angular resolution over which it is averaged. 
The bottom horizontal axis is the angular resolution, the top vertical one is 
the corresponding impact parameter $b_{\rm res}=D\sin(\theta_{\rm res})$. 
The left vertical axis is $2r_s{\cal J}^{n,\gamma}$ --- see 
Sec.~\ref{subsec:los_ave} --- while the right vertical axis translates those 
values in terms of optical depth for typical DM parameters, but only for 
annihilating DM (the optical depth axis for decaying DM would be a factor 
of $\sim 1.5$ smaller, for nominal parameters --- see Eq.~\ref{eq:dm_numbers}).}
\label{fig:res}
%\end{center}
}%\end{figure}

\subsubsection{The line-of-sight approximation}
\label{subsubsec:los_approx}
With the same approximation as above (see Eq.~\ref{eq:approx}), one can 
also try to find analytical solutions to the line-of-sight integral 
of Eq.~(\ref{eq:Jlos}). Assuming that the impact parameter $b\ll r_s \ll 
\ll D$, which is always the case for typical angular 
resolutions, we can write ${\cal J}^{n, \gamma} \approx 
{\cal J}^{n\gamma}_{\rm los}$, where:
\ben
{\cal J}^{n\gamma}_{\rm los} &\equiv& \frac{(1-\mu_{\rm res})}{b_{\rm res}^2} 
\int_0^{b_{\rm res}} db\,b\,\int_0^{r_s}
 \frac{ds}{r_s}\left( \frac{r_s^2}{s^2+b^2}\right)^{n\gamma/2}\;.
\een
The upper bound of the integral over $s$ is $r_s$ and not $R_{\rm vir}$ to 
be consistent with the validity domain of our current approximations.
The different possible combinations of $n\gamma$ are recalled in 
Tab.~\ref{tab:ngamma}.

\TABLE{%\begin{table}
\centering
\begin{tabular}{cccc}
\hline
nature / profile & cored & NFW & Moore \\
& ($\gamma=0$) & ($\gamma=1$) & ($\gamma=3/2$) \\
decaying ($n=1$) & $0$  & $1$ & $3/2$ \\
annihilating ($n=2$) & $0$ &$2$ & $3$ \\
\hline
\end{tabular}
\caption{Various combinations of the product $n\gamma$ and associated DM 
configurations for the analytical functions ${\cal J}^{n\gamma}_{\rm sph.}$ 
and ${\cal J}^{n\gamma}_{\rm los}$ (see Secs.~\ref{subsubsec:sph_approx} 
and~\ref{subsubsec:los_approx}).}
\label{tab:ngamma}
}%\end{table}

The function ${\cal J}^{n\gamma}_{\rm los}$ can be computed analytically in the 
following cases:
\ben
\label{eq:los_approx}
{\cal J}^0_{\rm los} &=& \frac{(1+\mu_{\rm res})}{2}\\ 
{\cal J}^1_{\rm los} &=& \frac{(1+\mu_{\rm res})}{2} 
\ln\left(\frac{2\,r_s}{b_{\rm res}}\right) \nn\\
{\cal J}^2_{\rm los} &=& \frac{(1+\mu_{\rm res})\pi}{2} 
\frac{r_s}{b_{\rm res}}\nn\\
{\cal J}^3_{\rm los} &=& (1+\mu_{\rm res}) \left(\frac{r_s}{b_{\rm res}}\right)^2 
\ln\left( \frac{b_{\rm res}}{(1+\sqrt{2})r_{\rm cut}} \right)\;.\nn
\een
Armed with these equations, the approximated optical depths are merely 
proportional to $2\,r_s\,{\cal J}^{n\gamma}_{\rm los}$, as given in 
Eqs.~(\ref{eq:tau_dm_av}) and (\ref{eq:av_nel_exact}). The results are also 
reported in Fig.~\ref{fig:res}, where the agreement with the exact calculation 
is shown to be quite good.

\subsection{Thermal to DM optical depth ratio $\eta_\chi$}
\label{subsec:eta_dm}

To size the relative amplitude of the SZ distortion induced by DM annihilation 
or decay on top of the thermal component, it is useful to derive the ratio 
$\eta_\chi \equiv \langle \tau_{e,\chi} \rangle_{\rm res}/
\langle \tau_{e,{\rm th}} \rangle_{\rm res} \gtrsim 0.1$, at any frequency 
where $\Delta I_{\rm th} \neq 0$. Since the frequency where 
$\Delta I$ crosses 0 is different for each electron component, it is further 
important to compare the absolute amplitude of $\Delta I_{\chi}$, at the 
frequency where $\Delta I_{\rm th} = 0$, with the current experimental 
sensitivities, \ie\ $\Delta I/ I \sim 10^{-6}$. We will 
estimate the ratio $\eta_\chi$ below, but postpone our discussion of the latter 
point in Sec.~\ref{sec:res}. For the sake of clarity, we will base the 
following calculation on the assumption that the DM density profile is an NFW 
($\gamma =1$). Likewise, we will use the analytical line-of-sight 
approximation derived in the previous section, which turns out to be the most 
precise.

%\begin{widetext}

Using Eqs.~(\ref{eq:dm_numbers}) and~(\ref{eq:th_numbers}), the ratio of the 
DM to thermal optical depths is proportional to 
${\cal J}_{\rm los}^{1(2)}/{\cal J}_{\rm los}^0$, 
for decaying (annihilating) DM. More explicitly, we get, for decaying DM,
\ben
\eta_\chi^{\rm dec} &\equiv& 
\frac{ \langle \tau_{e,\chi}^{1,1} \rangle_{\rm res}^{\rm dec} }
{\langle \tau_{e,{\rm th}} \rangle_{\rm res} } \nn \\
&=& 
\frac{\tau_{0,\chi}^{\rm dec}}{\tau_{0,{\rm th}}}
\, \left[ \frac{n_{e,{\rm th}}^0}{0.01\,{\rm cm^{-3}}}\right]^{-1} 
\, \frac{N_0}{10} \, \frac{\bar{\cal F}}{10^3} 
\, \frac{\Gamma_\chi }{10^{-26}{\rm s^{-1}}} 
\, \frac{\tau_{\rm loss}}{10^{17}{\rm s}} 
\, \left[ \frac{\rho_s /(0.05\,{\rm GeV/cm^3})}{\mchi /{\rm GeV}}\right]
\, \left[ \frac{{\cal J}_{\rm los}^1}{{\cal J}_{\rm los}^0} \right]\nn\\
&=& \eta_{\chi}^{0,{\rm dec}} 
\, \left[ \frac{n_{e,{\rm th}}^0}{0.01\,{\rm cm^{-3}}}\right]^{-1} 
\, \frac{N_0}{10} \, \frac{\bar{\cal F}}{10^3} 
\, \frac{\Gamma_\chi }{10^{-26}{\rm s^{-1}}} 
\, \frac{\tau_{\rm loss}}{10^{17}{\rm s}} 
\, \left[ \frac{\rho_s /(0.05\,{\rm GeV/cm^3})}{\mchi /{\rm GeV}}\right]
\times \nn\\
&& \ln\left[\frac{2 (r_s/400 \,{\rm kpc})}{b_{\rm res}/1 \,{\rm kpc}}\right]\;,
\label{eq:eta_approx_dec}
\een
with 
\ben
\eta_{\chi}^{0,{\rm dec}} = 3.29\times 10^{-4}\;.
\een
For annihilating DM, the result is proportional to 
${\cal J}_{\rm los}^2/{\cal J}_{\rm los}^0$, and we have instead 
\ben
\eta_\chi^{\rm ann} &\equiv& 
\frac{\langle \tau_{e,\chi}^{1,1} \rangle_{\rm res}^{\rm ann}}
{\langle \tau_{e,{\rm th}} \rangle_{\rm res}}\nn\\ 
&=& \eta_{\chi}^{0,{\rm ann}} 
\, \left[ \frac{n_{e,{\rm th}}^0}{0.01\,{\rm cm^{-3}}}\right]^{-1} 
\, \frac{N_0}{10} \, \frac{\bar{\cal F}}{10^3} 
\, \frac{\sigv }{3\cdot 10^{-26}{\rm cm^{3}s^{-1}}} 
\, \frac{\tau_{\rm loss}}{10^{17}{\rm s}} 
\, \left[ \frac{\rho_s /(0.05\,{\rm GeV/cm^3})}{\mchi /{\rm GeV}}\right]^2
\times \nn\\
&& \left[\frac{r_s/400 \,{\rm kpc}}{b_{\rm res}/1 \,{\rm kpc}}\right]\;,
\label{eq:eta_approx_ann}
\een
where
\ben
\eta_{\chi}^{0,{\rm ann}} = 4.71\times 10^{-3}\;.
\een

%\end{widetext}

The relevant parameters here are $\eta_{\chi}^{0,{\rm dec}}$ and 
$\eta_{\chi}^{0,{\rm ann}}$, which both are $\ll 0.1$ for most of DM models and 
for current and future experimental performances. Because the optical depth 
sizes the amplitude of the spectral distortion~(see Eq.~\ref{eq:deltaI}),
this means that detecting a SZ signal from DM annihilation or decay on top of 
the thermal contribution demands very strong spectral distortions of the 
former with respect to the latter. We study the spectral features of the SZ 
signatures in the next section.

\section{Spectral distortion analysis}
\label{sec:res}

To complete our study, we have to tackle a full spectral analysis. 
Nevertheless, what is quantitatively interesting is the amplitude 
of the spectral change due to DM at the SZ transition frequency of the thermal
gas, \textit{i.e.} when its global effect is null. In this section, therefore, 
we estimate the amplitude of $\Delta I_{\chi}$ at the frequency for which 
$\Delta I_{\rm th} = 0$, to size the potential impact of DM at the transition 
frequency of the thermal background. We adopt the formalism 
developed in~\cite{2000A&A...360..417E}, which is very well suited for our 
calculation. We refer the reader to that article for more details on the 
quantities and results that are used below.

Let us first define the reduced blackbody intensity 
$i(x)\equiv I(E_k)/i_0 = x^3/(e^x-1)$, where $x\equiv E_k/(kT_0)$, $T_0$ is 
the CMB temperature, and $i_0\equiv 2(kT_0)^3/(hc)^2$. With these notations, 
Eq.~(\ref{eq:deltaI}) can be rewritten in terms of the reduced intensities
\ben
\delta i(x) = \tau \, (j(x) - i(x))\;,
\label{eq:deltai}
\een
where we further define the scattered photon intensity as:
\ben
j(x) \equiv \int_{0}^{t_{\rm max}} dt \, P_1(t)\,i(x/t)\;.
\label{eq:jx}
\een
The photon frequency redistribution function $P_1(t)$, valid in the single 
scattering approximation, depends on the normalized injected electron spectrum 
${\cal F}(E)={\cal F}(\bar{p})$  
($\bar{p}\equiv p/(m_ec)=\gamma_e\beta_e=\sqrt{\gamma_{e}^2-1}$ being the 
reduced electron momentum) as follows:
\ben
P_1(t) = \int d\bar{p} \,{\cal F}(\bar{p})\,P_1(t,\bar{p})\;,
\label{eq:redistrib_func}
\een
where $P_1(t,\bar{p})$ has the following analytical form
\begin{align}
P_1(t,\bar{p})& = -\frac{3|1-t|}{32\bar{p}^{6} t}
\left\{ 1 + (10+8\bar{p}^2+4\bar{p}^4)t+t^2\right\} + \\
 \frac{3(1+t)}{8\bar{p}^5} &
\left\{ \frac{3+3\bar{p}^2+\bar{p}^4}{\sqrt{1+\bar{p}^2}} - 
\frac{3+2\bar{p}^2}{\bar{p}} 
\left({\rm asinh}(\bar{p}) -\frac{|\ln(t)|}{2} \right)  \right\}\;.
\nn
\end{align}
The maximal frequency shift, above which $P_1(t,\bar{p})=0$, fulfills the 
condition $|\ln(t_{\rm max})| = 2\, {\rm asinh}(\bar{p})$.

Armed with these equations, we can derive the intensity shift $\delta i $ 
for any electron component, given its spectrum. For the thermal electrons, we 
will use a Maxwell-Boltzmann spectrum
\ben
{\cal F}_{\rm th}(\bar{p}) = 
\frac{\beta_{\rm th}}{K_2(\beta_{\rm th})}\bar{p}^2\,
\exp\left(-\beta_{\rm th}\sqrt{1+\bar{p}^2}\right)\;,
\een
where $\beta_{\rm th} = m_ec^2/(kT_{\rm th})$ is the inverse reduced electron
temperature, and where the normalization is ensured with the modified Bessel 
function of the second kind $K_2$. The resulting redistribution function 
(see Eq.~\ref{eq:redistrib_func}) is shown in Fig.~\ref{fig:redist_th}, for 
different temperatures.

\FIGURE[t]{%\begin{figure}[t]
%\begin{center}
\includegraphics[width=\columnwidth, clip]{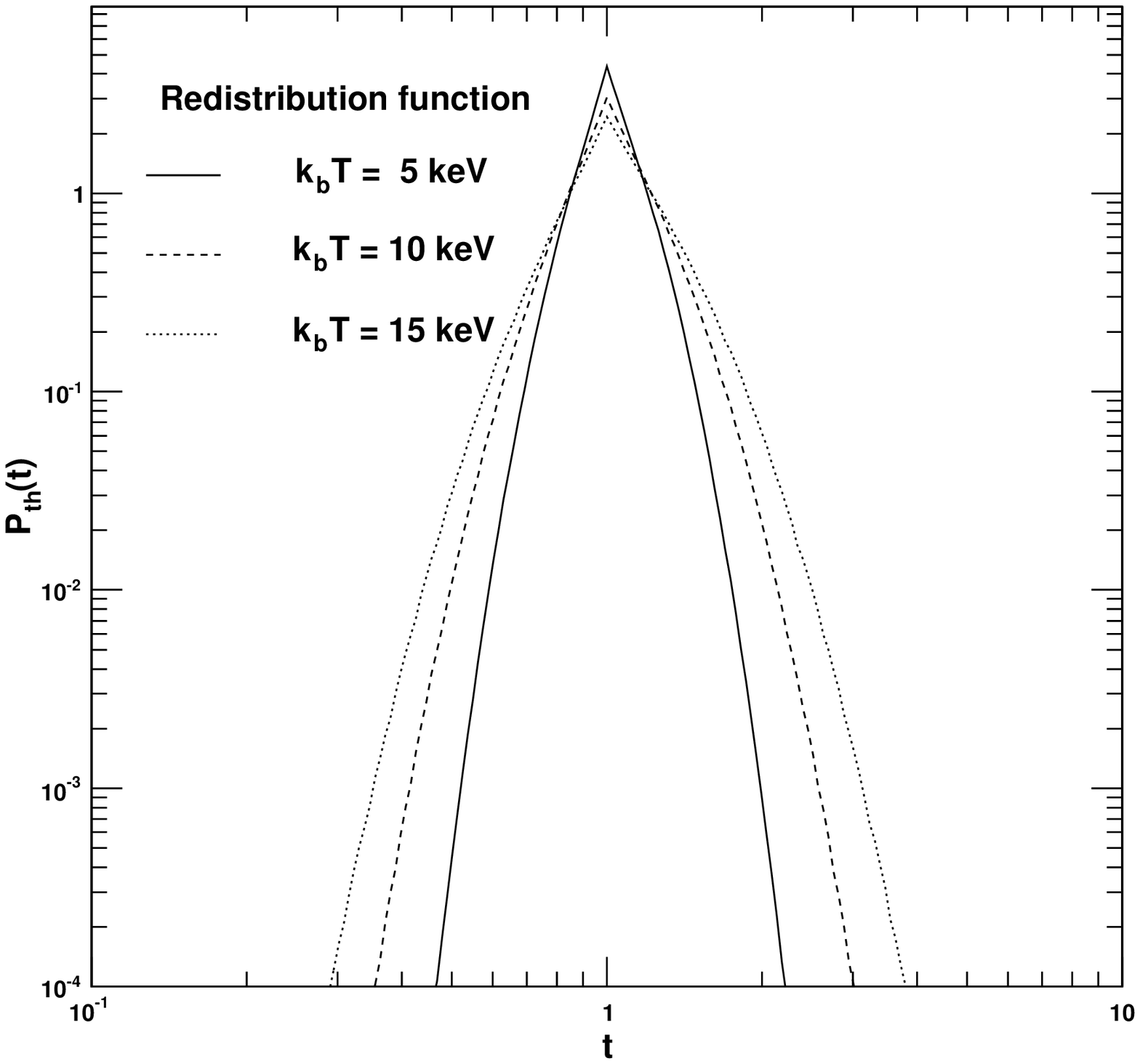}
\caption{Frequency redistribution function for different thermal electron gas 
temperatures, $kT_{\rm th} = 5,10,15$ keV.}
\label{fig:redist_th}
%\end{center}
}%\end{figure}

\FIGURE[t]{%\begin{figure}[t]
%\begin{center}
\includegraphics[width=\columnwidth, clip]{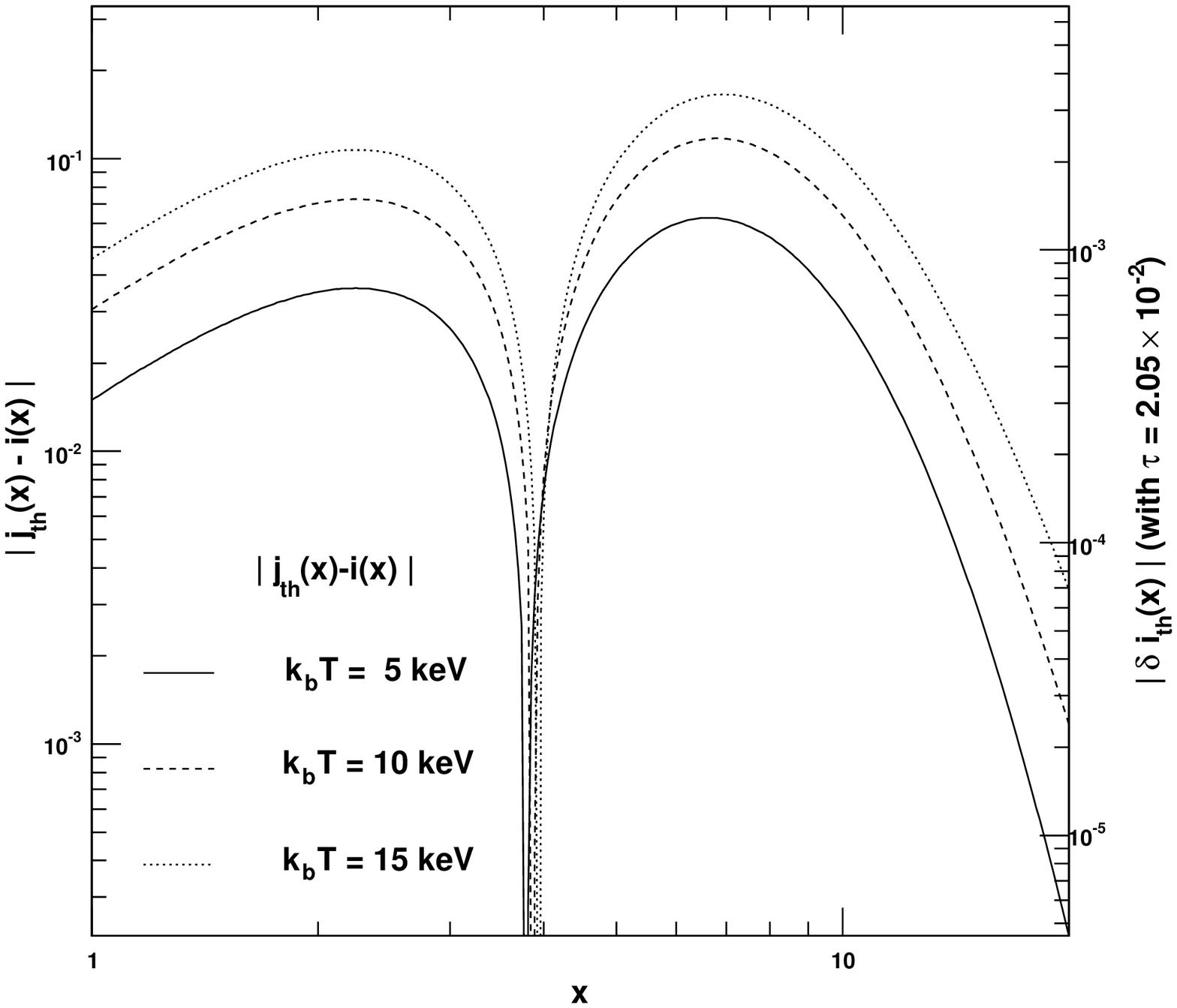}
\caption{$\delta i_{\rm th}(x)/\tau$ for different thermal electron gas 
temperatures, $kT_{\rm th} = 5,10,15$ keV (the right vertical axis gives 
$|\delta i_{\rm th}(x)|$ for our template cluster, with an optical 
depth of $\tau = 2.05\times 10^{-2}$).}
\label{fig:deltai_th}
%\end{center}
}%\end{figure}

For the DM electron yield, we will assume that the injected spectrum is 
$\propto \delta(E-n\,m_\chi/2)$, with $n = 1$ or 2 for decay or annihilation, 
respectively. This is equivalent to assuming a direct annihilation or decay in 
$e^+e^-$, which is particularly relevant for light DM particles but a very 
optimistic spectral hypothesis in the general case. We can thus determine the 
normalized equilibrium spectrum with Eq.~(\ref{eq:dnde_dm}):
\ben
{\cal F}(E) = \frac{K}{b(E)}\;\;\;\;\;({\rm for}\;\;E\leq n\,\mchi / 2)\;,
\label{eq:fe_dm}
\een
where $K=K(E_{\rm min},E_{\rm max})$ is a normalization constant
such that $\int dE {\cal F}(E) = 1$ in the energy range of interest. For the 
energy loss rate $b(E)$, we can combine the two most relevant regimes, 
\emph{i.e.} the Coulomb losses below $E_0=1$ GeV, of timescale $\tau_{\rm Coul.}$
and the inverse Compton losses on CMB above~\cite{1999ApJ...520..529S}, of 
timescale $\tau_{\rm IC}$. Actually, we have:
\ben
\frac{b(E)}{10^{-16}{\rm GeV/s}} &=& b_{\rm IC}(E)+ b_{\rm Coul.}(E)
\label{eq:eloss}\\
&=&  0.265\left(\frac{E}{E_0}\right)^2 +
\frac{6.20}{\beta_e}\frac{n_{e,{\rm th}}}{1\,{\rm cm^{-3}}}
\left\{ 1+ \ln\left( \frac{E/E_0}{n_{e,{\rm th}}/1\,{\rm cm^{-3}}} \right)
\right\}\;,\nn
\een
where the Coulomb loss rate is taken from~\cite{1979tpa..book.....G}. We can
simplify the previous expression by means of the timescales
\ben
b(E) &\simeq& \frac{E_0 }{\tau_{\rm Coul.}}+\frac{E^2}{E_0\tau_{\rm IC}}
= \frac{E_0}{\tau_{\rm Coul.}}
\left\{ \alpha_\tau \left(\frac{E}{E_0}\right)^2 + 1\right\}\nn\\
\frac{\tau_{\rm Coul.}}{10^{16}\,{\rm s}} &\equiv& 0.16\times 
\frac{\beta_e}{n_{e,{\rm th}}/1\,{\rm cm^{-3}}}\;;
\;\;\;\frac{\tau_{\rm IC}}{10^{16}\,{\rm s}} \equiv 3.77 \;\;\;{\rm and} \;\;\; 
\alpha_\tau \equiv    \frac{\tau_{\rm Coul.}}{\tau_{\rm IC}}\;.
\label{eq:eloss_approx}
\een
With this form of energy loss rate, we can reexpress Eq.~(\ref{eq:fe_dm}) as
\ben
{\cal F}(E) =
\frac{K(E_{\rm min},\mchi)\,\tau_{\rm coul}}{\alpha_\tau \frac{E^2}{E_0^2} + 1}
\;.
\een
The correspondence between energy and reduced momentum is straightforward: 
$\bar{p} = \sqrt{\gamma_e^2-1}$, and $dE = \bar{p}m_e^2c^4 d\bar{p}/E$. The 
energy lower bound $E_{\rm min}$ must be consistent with the cluster age, 
DM-induced electrons having essentially been injected since the cluster has 
formed. Therefore, these electrons had only a time of the order of the cluster 
age $\Delta t \approx t_{\rm cl} $ to lose their energy. We can compute the 
minimal energy $E_{\rm min}\geq m_ec^2$ corresponding to any injected energy $E$ 
by demanding $\int_{E_{\rm min}}^{E}dE'\,\frac{E_0}{b(E')} \leq \Delta t$, which is
equivalent to $K^{-1}(E_{\rm min},E) \leq \Delta t$. For typical cluster 
formation redshifts around 5, $E_{\rm min}$ is merely found to be $m_ec^2$.

The photon redistribution function obtained from the DM-induced electrons, 
once the normalized spectrum defined in Eq.~(\ref{eq:fe_dm}) has been injected 
into Eq.~(\ref{eq:redistrib_func}), is shown in Fig.~\ref{fig:redist_dm}, for 
different WIMP masses, assuming direct annihilation or decay in $e^+e^-$.

\FIGURE[t]{%\begin{figure}[t]
%\begin{center}
\includegraphics[width=\columnwidth, clip]{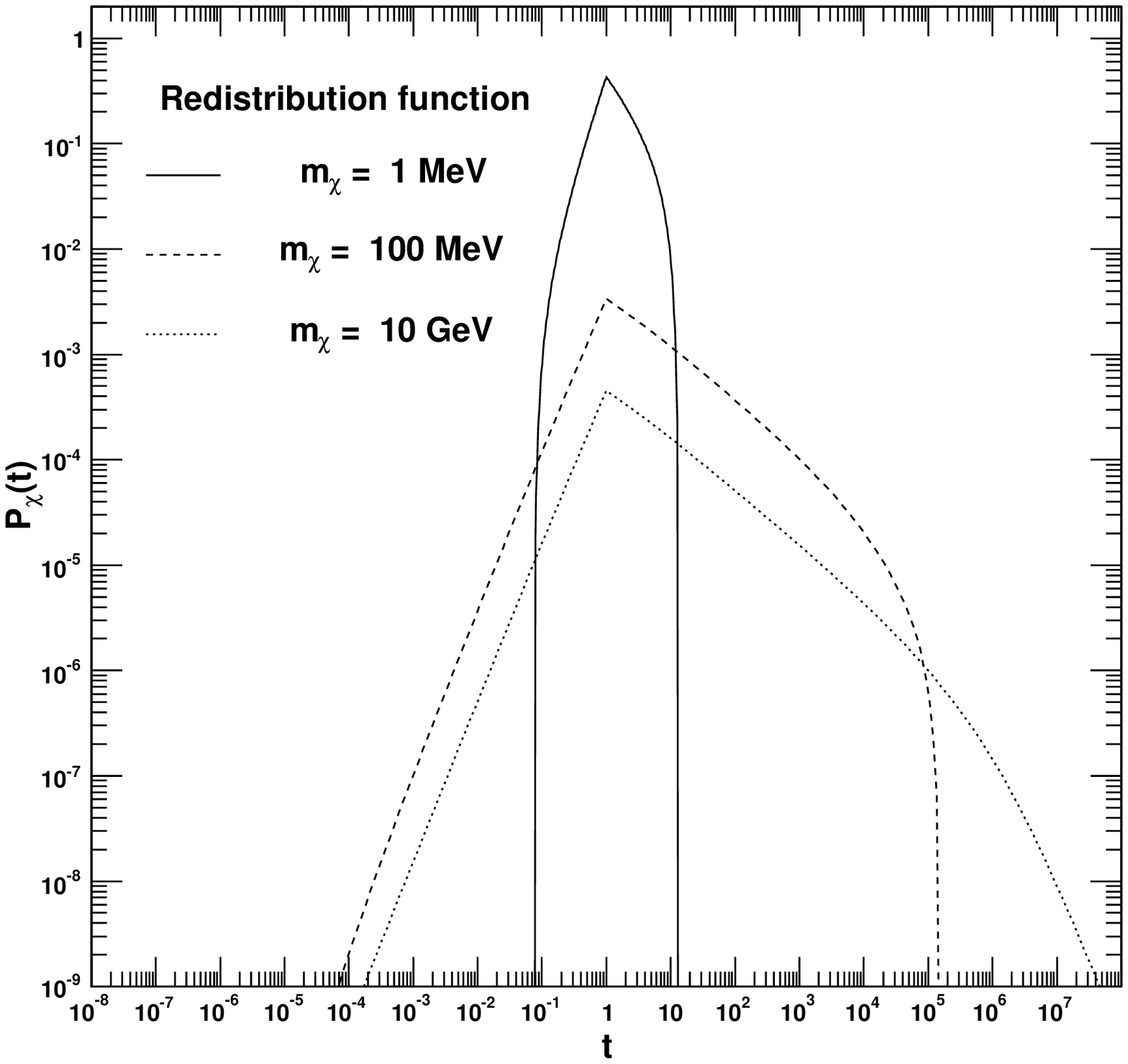}
\caption{Redistribution functions for different WIMP masses, assuming 
$\chi\chi\rightarrow e^+e^-$. Decaying DM functions are identical for 
$\chi\rightarrow e^+e^-$ but with twice the indicated masses.}
\label{fig:redist_dm}
%\end{center}
}%\end{figure}

Notice that when coming to the numerical calculation of the optical 
depth, we will take advantage of the estimates performed in the previous 
section by remarking that the function $\bar{\cal F}$ defined in 
Eq.~(\ref{eq:def_fbar}) obeys rigorously $\tau_{\rm loss}\bar{\cal F} = 
K^{-1}(E_{\rm min},\mchi)$, in the current DM configuration. If we assume that 
$\tau_{\rm loss}\approx \tau_{\rm coul} \approx \tau_{\rm ic}$, then we have 
$\bar{\cal F} \approx 1$. Hence, with such an assumption, we can rescale 
the optical depths given in Eq.~(\ref{eq:typ_tau_dm}) accordingly.

Our aim is to size $\delta i_\chi(x_{\rm th})$ at a frequency $x_{\rm th}$ 
such that $\delta i_{\rm th}(x_{\rm th})=0$. $x_{\rm th}$ is readily found by 
computing Eq.~(\ref{eq:deltai}), given the thermal electron temperature. 
For completeness, we take three different temperatures, 
$kT_{\rm th} =5,10,15$ keV, though it is well known that the transition 
frequency is much less dependent on the temperature than the amplitude, as 
shown in Fig.~\ref{fig:deltai_th}. As illustrated in that plot for the three 
cases, we find $x_{\rm th} \simeq  4$. We can now calculate
\ben
\delta i_\chi(x_{\rm th}) = \tau_\chi (j_\chi(x_{\rm th})-i(x_{\rm th}))\;,
\een
with
\ben
j_\chi(x_{\rm th}) = \int_{0}^{t_{\rm max}} dt \, P_1(t,\bar{p}_\chi)\,i(x/t)\;,
\een
and for different DM particle masses.

We first report the behavior obtained for $\delta i_\chi(x)/\tau_\chi$ with 
different WIMP masses in Fig.~\ref{fig:dix_mchi}, assuming direct annihilation
in $e^+e^-$ (also valid for decaying DM, but corresponding then to twice 
heavier WIMP masses). 
The factor of $1/\tau_\chi$ allows a prediction independent of the electron 
density, \ie\ independent of the cluster halo parameters. The 
right vertical axis gives the conversion in terms of $\delta i_\chi(x)$, 
assuming $\tau_{\chi} = 1.5\times 10^{-9}$ (see below), which reaches a maximum 
of $\delta i\sim 10^{-10}$ around the transition frequency of the thermal 
component $x_{\rm th}\sim 4$. This maximal value corresponds to 
an annihilating DM of 1 GeV distributed with an NFW profile and observed 
from our template cluster within an angular resolution of $10''$ 
(see Eq.~\ref{eq:dm_numbers} and Fig.~\ref{fig:res}). It is anyway easily 
rescaled for other configurations.

In Fig.~\ref{fig:di_mchi}, we also trace 
$\delta i_\chi(x_{\rm th})/(\tau_\chi m_\chi^n i_0(x_{\rm th}))$, \ie\ at the 
transition frequency of the thermal component, as a function of the DM mass 
$\mchi$; $n=$1 (2) for decaying (annihilating) DM. We implement the expected 
mass dependence of the optical depth thanks to the factor of $1/m_\chi^n$. To 
compare with experimental sensitivities, we have converted our result in terms 
of the relative intensity shift $\delta i$, but only in the case of DM 
annihilation (\cf~the right vertical axis). To do so, we have used the 
angular average of $\tau_\chi$ as derived in Eq.~(\ref{eq:dm_numbers}), which 
rests on the template DM halo defined in Tab.~\ref{tab:cl}. Likewise, to be 
consistent with the assumption of annihilation or decay in $e^+e^-$, however, we
have used $N_0 = 2$, and, as discussed above, $\bar{\cal F} = 1$ (\ie\ for 
$\tau_{0,\chi}$, we have taken the values of Eq.~\ref{eq:typ_tau_dm} divided 
by $5\cdot 10^3$, which gives $\tau_{0,\chi} = 1.5\times 10^{-9}$). To switch 
to the $\delta i $ associated with DM decay, one may apply a factor of 
$1.01/7.68 \approx 0.13 $ to the right vertical scale (see 
Eq.~\ref{eq:dm_numbers}). Notice that the scaling with $\mchi$ is 
trivial, $\propto 1/\mchi^{2}$ for DM annihilation, and $\propto 1/\mchi$ for 
DM decay.

\FIGURE[t]{%\begin{figure}[t]
%\begin{center}
\includegraphics[width=\columnwidth, clip]{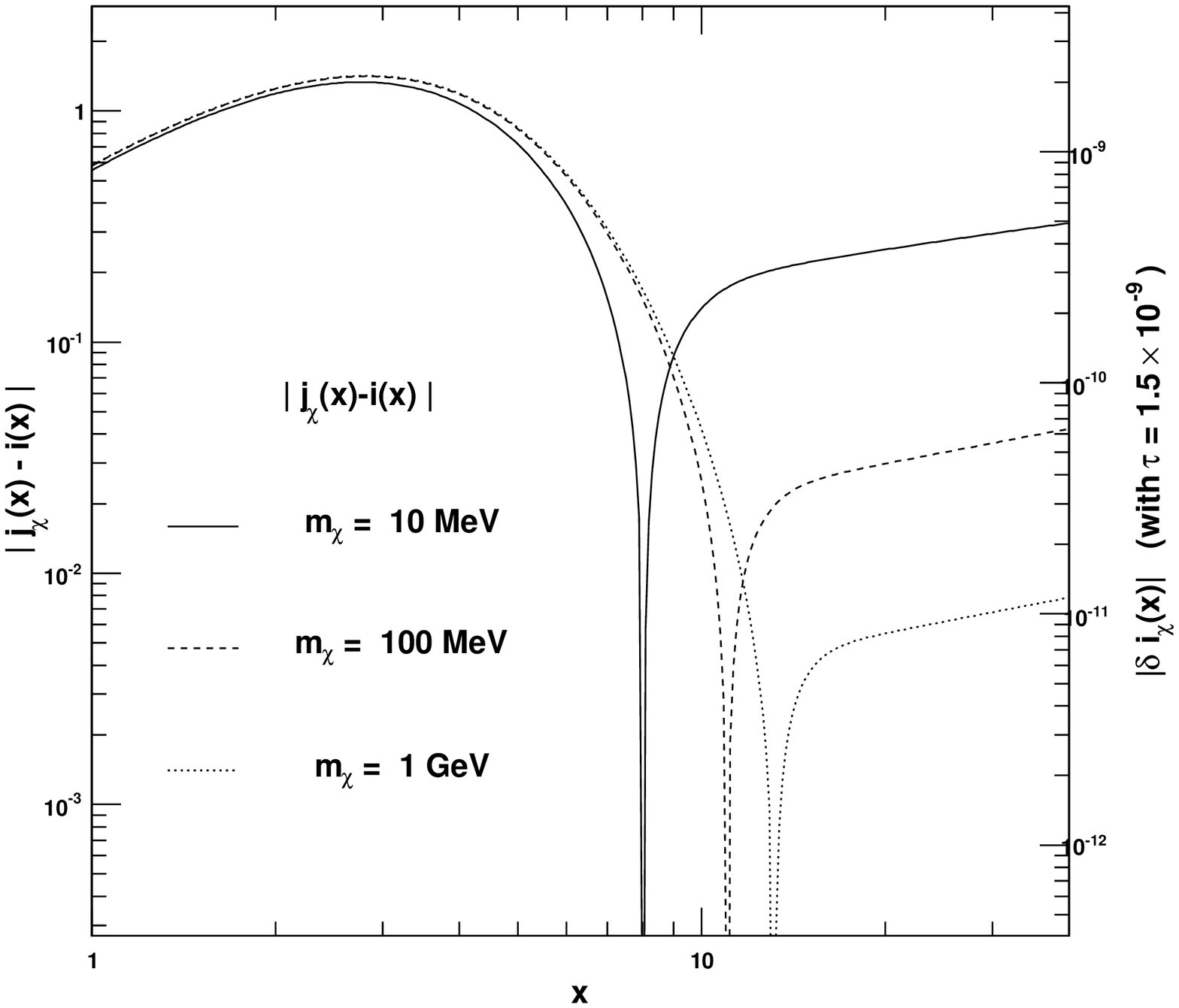}
\caption{$|\delta i_\chi(x)|/\tau_\chi=|j_\chi(x)-i(x)|$ as a function of the 
reduced frequency $x$ for different WIMP masses, assuming direct annihilation 
in $e^+e^-$ (identical for decaying DM, but corresponding to twice the masses). 
The right vertical axis provides the conversion to $|\delta i_\chi(x)|$ for 
an optical depth of $\tau = 1.5\times 10^{-9}$, corresponding to a 
configuration with 1 GeV annihilating DM with an NFW profile inside our 
template cluster observed with an angular resolution of $10''$ --- see 
Eq.~(\ref{eq:dm_numbers}) and Fig.~\ref{fig:res}.}
\label{fig:dix_mchi}
%\end{center}
}%\end{figure}

\FIGURE[t]{%\begin{figure}[t]
%\begin{center}
\includegraphics[width=\columnwidth, clip]{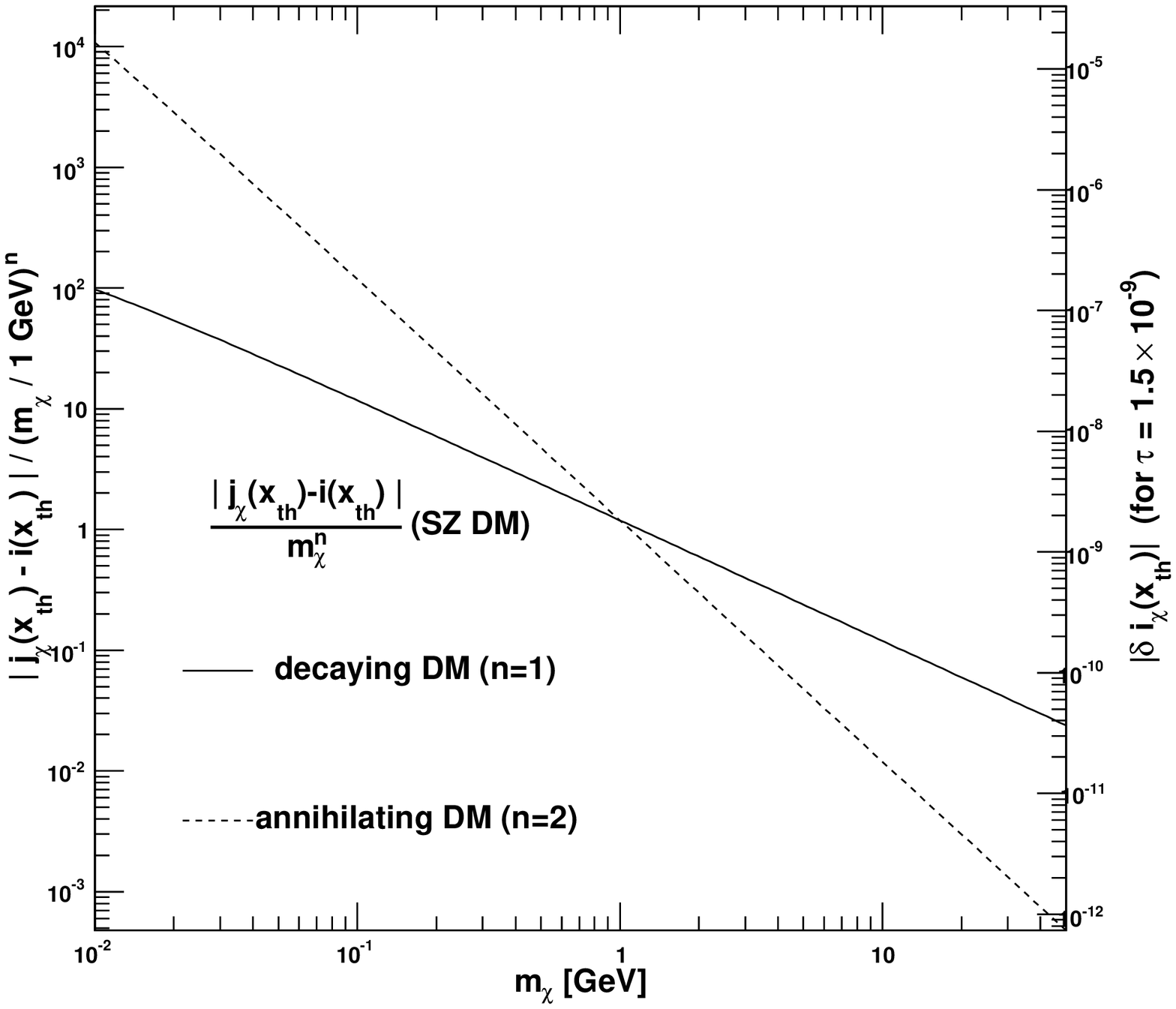}
\caption{$\delta i_\chi(x_{\rm th})/(m_\chi^n \tau_\chi)$ as a function of the DM 
particle mass $\mchi$, for $x_{\rm th}=4$ (left vertical axis). The right 
vertical axis provides the conversion to $|\delta i_\chi(x_{\rm th})|$ for 
an optical depth of $\tau = 1.5\times 10^{-9}$, corresponding to a 
configuration with annihilating DM with an NFW profile inside our template 
cluster observed with an angular resolution of $10''$ --- see 
Eq.~(\ref{eq:dm_numbers}) and Fig.~\ref{fig:res}.}
\label{fig:di_mchi}
%\end{center}
}%\end{figure}

From the results presented in Figs.~\ref{fig:dix_mchi} and~\ref{fig:di_mchi}, 
it seems quite difficult to detect any SZ signal from DM annihilation or 
decay, since the typical values obtained for the relative intensity shift 
is $\delta i_\chi \sim 10^{-10}$, still far from current experimental 
sensitivities (Planck was optimized for 
$\delta i \sim 10^{-6}$~\cite{2006astro.ph..4069T}, \ie\ the $\mu$K level in 
terms of temperature). However, we remind that this particular value of the 
intensity shift is connected to some assumptions. For the line-of-sight 
integral, which depends on the cluster halo properties, we have used 
$2 r_s {\cal J} = 10^3$ and $10^5$ kpc for DM decay and annihilation, 
respectively, which corresponds to observing our NFW template cluster within 
an angular resolution of $\sim 10''$ (see Fig.~\ref{fig:res}). Moreover, 
there are additional assumptions on $\mchi \sim 1$ GeV and on the DM 
properties, though quite generic (see Eq.~\ref{eq:dm_numbers}). Anyway, 
we emphasize that these assumptions are rather optimistic as regards
the DM modeling, and should therefore be considered as conservative. For 
comparison, the intensity shift is shown to be much larger for thermal 
electrons (see Fig.~\ref{fig:deltai_th} with the vertical right axis), 
$\delta i \sim 10^{-4}-10^{-3}$, over a broad band of frequencies, except at 
the transition frequency. Hence, the DM effect is likely too small in the 
major part of the ``natural'' DM parameter space, but might be still sizable 
for very light DM candidates and very cuspy profiles.

To figure out the influence of both the DM halo cuspiness and the experimental 
angular resolution, one can read off the left vertical axis of 
Fig.~\ref{fig:res}, which provides values of the line-of-sight integral for 
different cluster halo profiles as functions of the angular resolution. 
To boost our predictions by $\gtrsim$ 5 orders of magnitude we would have to
consider very cuspy profiles ($\gamma \gtrsim 1.5$) and in the meantime very 
deep spatial resolutions ($b_{\rm res} \sim 0.1$ kpc, or equivalently
$\theta_{\rm res}\sim 1''$), at least for the template 
set of halo parameters of Tab.~\ref{tab:cl}. We will show in 
Sec.~\ref{sec:diff}, however, that even this over-optimistic case is actually
limited by spatial diffusion effects.

If we stick to an NFW profile, an alternative to enhance our predictions would 
be to consider very light annihilating WIMPs, in the sub-GeV mass-scale. 
However, in that case, the value of the annihilation cross considered in 
Eq.~(\ref{eq:dm_numbers}) is bounded by astrophysical observations in the 
Milky Way. Indeed, it was found 
in~\cite{2004PhRvL..92j1301B,2006MNRAS.368.1695A}, that the annihilation cross 
section for light DM must obey $\sigv\lesssim 10^{-31} (\mchi/{\rm MeV})^2 \,
{\rm cm^3s^{-1}}$ to not overshoot the 511 keV bulge emission measured by the 
SPI spectrometer onboard the INTEGRAL satellite \cite{2005A&A...441..513K}.
Since Fig.~\ref{fig:di_mchi} was done with a cross section of 
$\sigv = 3\cdot 10^{-26} {\rm cm^3s^{-1}}$, such a limit translates into
a plateau below $\mchi \sim 500$ MeV, translating into a saturation of 
$\delta i_\chi\lesssim 10^{-8}$, much below the current and future 
sensitivities.

Invoking very cuspy profile with $\gamma >1 $ appears be the only possibility 
left for DM to generate an observable SZ distortion of the CMB spectrum, but 
such an extreme configuration (i) is hardly motivated from theoretical and 
observational constraints and (ii) will even be shown, in the
next section, to have limited effect due to spatial diffusion. Last but not 
least, one should not forget about the additional foreground coming from other 
relativistic electrons also injected in clusters from standard astrophysical 
sources \cite{2000A&A...360..417E}. In any case, any attempt of DM 
interpretation of any SZ imprints would seem daring, at least in our view.

\section{Impact of spatial diffusion: analytical insights}
\label{sec:diff}

So far, we have assumed that the equilibrium electron density 
was set only from energy losses and neglected the spatial transport, which
was presented as a reasonable hypothesis in~\cite{2006A&A...455...21C} 
for angular resolutions $\lesssim 1'$. Nevertheless, as discussed above, 
angular resolutions down to $1''$, as expected in future experiments, 
correspond to quite small spatial scales for typical nearby clusters, so that 
the relevance of such an assumption can be questioned.

If we include spatial diffusion, then Eq.~(\ref{eq:diff_eq}) must be rewritten
in terms of a current conservation equation~\cite{1964ocr..book.....G}
\ben
-\vec{\nabla}\left(K_r \vec{\nabla} \frac{dn}{dE}\right) - 
\partial_E  \left( b(E) \frac{dn}{dE} \right) = {\cal Q}(\vec{x},E)\;.
\label{eq:diff_eq2}
\een
$K_r$ is the diffusion coefficient that characterizes the stochastic transport 
of electrons caused by their diffusion on the magnetic inhomogeneities of the 
intracluster medium. We will assume this coefficient to depend on the energy 
only $K_r \equiv K_r(E)$. This equation can be solved by looking for the 
associated Green function, which is readily found in infinite 3D space:
\ben
{\cal G}(\vec{x},E\leftarrow \vec{x}_s,E_s) = 
\frac{1}{b(E) \pi^{\frac{3}{2}}\lambda^3} 
\exp\left\{ - \frac{(\vec{x}_s-\vec{x})^2}{\lambda^2} \right\}\;,
\label{eq:g3D}
\een
where $\lambda$ is a characteristic propagation scale defined as
\ben
\lambda^2(E,E_s) \equiv 4 \int_E^{E_s}dE' \frac{K(E')}{b(E')}\;.
\label{eq:def_lambda}
\een
The electron density at any position $\vec{x}$ and energy $E$ in the cluster 
is then given by 
\ben
\label{eq:exact_ne}
\frac{dn}{dE}(\vec{x},E) &=& \int_E^{n\,m_\chi/2} dE_s \int d^3\vec{x}_s \,
{\cal G}(\vec{x},E\leftarrow \vec{x}_s,E_s)\,{\cal Q}(\vec{x}_s,E_s)\;.
\een
We can remark that we recover the diffusion-less case in the limit of 
vanishingly small propagation scale $\lambda\rightarrow 0$, since in this case 
we have ${\cal G}(E,\vec{x}\leftarrow \vec{x}_s)\leftrightarrow\delta^3
(\vec{x}-\vec{x}_s)/b(E)$. For a monochromatic injection of electrons from DM 
annihilation or decay, the diffusion-less limit is therefore recovered when 
$E\rightarrow n\,m_\chi/2$. More generally however, since the Green function 
exhibits a Gaussian behavior of typical scale $\lambda$, the electron density 
is expected to be smeared out over this scale, which we actually demonstrate 
below.

The Green function appearing in Eq.~(\ref{eq:g3D}) was derived assuming 
diffusion in an infinite 3D space, which is obviously not the case since 
clusters have finite sizes. This still holds provided the propagation scale 
$\lambda$ is much lower than the typical size of the cluster, and while 
electron injection at the border of the object is irrelevant compared to 
injection in the more central regions.
If set by magnetic inhomogeneities, the diffusion coefficient can be expressed
as a function of the regular component of the magnetic field. Although 
it is absolutely not clear whether or not a Kolmogorov spectrum is relevant 
to describe magnetic turbulence in clusters, since it is not the case at the 
Galactic scale, it is still conventional to adopt such a behavior for the 
diffusion coefficient \cite{1999APh....12..169B}:
\ben
K_r(E) = K_0 \,\left[ \frac{l_B}{20\,{\rm kpc}} \right]^{2/3} \, 
\left[ \frac{B}{\mu{\rm G}} \right]^{-1/3} \,
\left[ \frac{E}{\rm GeV} \right]^{1/3}\;,
\label{eq:diff_coef}
\een
where $B$ is the magnetic field and $l_B$ its coherence length. With the above 
values, which we will use in the following, the normalization of the diffusion 
coefficient can set to 
$K_0 \sim 2.3\times 10^{29}{\rm cm^2s^{-1}}$~\cite{1999APh....12..169B}. Since 
the typical energy loss timescale is of order $\tau_{\rm loss}\sim 10^{16}
{\rm s}$, we have $\lambda \sim 2\sqrt{K_0\,\tau_{\rm loss}} \sim 30\, 
{\rm kpc}\ll r_s$, which is actually quite large though still smaller than 
the typical scale radii of clusters --- note that this diffusion scale will
slightly decrease with energy above 1 GeV, when the inverse Compton 
dominates over the Coulomb losses. Therefore, because the injection rate of 
electrons is expected to dominate within a volume set by $r_s\ll R_{\rm vir}$
and since $\lambda \ll r_s$, we can safely disregard the radial boundary 
condition and use the infinite 3D solution written above.

The source term ${\cal Q}$ is defined in Eq.~(\ref{eq:source_dm}). To simplify
the discussion, we will further suppose that DM annihilates or
decays into electron-positron pairs, so that $N_0=2$ and 
$F(E_s) = \delta(E_s-n\,m_\chi/2)$ --- $n=2$ in the case of annihilation, 1 in 
the case of decay. In this fiducial instance, the propagation scale only depends
on the energy $E$, $\lambda = \lambda(E,n\,m_\chi/2)$ --- 
see Eq.~(\ref{eq:def_lambda}). Therefore, the electron density at any cluster 
radius $r$ and energy $E\leq n\,m_\chi/2$ reduces to
\ben
\frac{dn}{dE}(r,E) &=& \frac{4\,\alpha_n\,\rho_s^n}{b(E)\,\sqrt{\pi}\,\lambda^3}
\int_{-1}^{1} d\mu'\int dr'\,r'^2\, 
 \exp\left\{-\frac{r'^2 + r^2 - 2\, r\, r'\, \mu'}{\lambda^2}\right\}\,
f_\gamma^n(r')\;,
\een
where we have taken advantage of the spherical symmetry of the source term, 
the radial dependence $f_\gamma(r')$ of which has been defined in 
Eq.~(\ref{eq:dm_prof}). The integral over the cosine $\mu'$ is straightforward, 
and we are left with:
\begin{align}
\frac{dn}{dE}(r,E) &=
\frac{2\,\alpha_n\,\rho_s^n}{b(E)\,\sqrt{\pi}\,\lambda\,r} \,
 \sum_{k=0}^{k=1}(-1)^k  \int dr'\,r' \, f_\gamma^n(r') \,
\exp\left\{- \frac{(r-(-1)^k \,r')^2}{\lambda^2}\right\}
\\
&\simeq 
\frac{2\,\alpha_n\,\rho_s^n\,r_s^{n\gamma}}{b(E)\,\sqrt{\pi}\,\lambda\,r} \,
\sum_{k=0}^{k=1}(-1)^k \int_{r-\lambda}^{r+\lambda} 
dr'\,|r'|^{1-n\gamma} \,
\exp\left\{- \frac{(r-(-1)^k \,r')^2}{\lambda^2}\right\}\;.\nn
\end{align}
In the last line, we have employed the approximation given in 
Eq.~(\ref{eq:approx}), where the DM profile is taken as a simple power law; 
indeed, we are mostly interested, again, in the SZ contribution from the 
densest central parts of the cluster. Moreover, we have restricted the 
integral range to $[r-\lambda;r+\lambda]$, accounting for the fact that 
electrons coming from distances above $\lambda$ are Gaussianly depleted, 
consistently with Eq.~(\ref{eq:g3D}); we remind that $\lambda \sim 10$ kpc. In 
the very central region of the cluster and within this range for $r'$, the 
arguments in the exponentials $(r-(-1)^k \,r')\lesssim \lambda$, so that we 
can expand the integral to get more insights on the physics at stake. To the 
first order, the result is analytical, and we find after integration:
\ben
\frac{dn}{dE}(r,E) \simeq 
\frac{8\,\alpha_n\, \rho_s^n\,r_s^{n\gamma}}{b(E)\,\sqrt{\pi}\,\lambda^3}\,
\frac{(r+\lambda)^{3-n\gamma}}{(3-n\gamma)}\,
\left[ 1 - {\rm sign}(r-\lambda)
 \frac{|r-\lambda|^{3-n\gamma}}{(r+\lambda)^{3-n\gamma}} \right] \;,\nn
\een
where we let the reader derive the logarithmic expression that arises when 
$3-n\gamma=0$. This result is very interesting, since it states that at the 
center of the cluster, \ie~$r=0$, the electron density is no longer diverging 
like $r^{-n\gamma}$ when diffusion occurs, but instead behaves like:
\ben
\frac{dn}{dE}(0,E) &\approx& 
\frac{16\,\alpha_n\, \rho_s^n(3-n\gamma)^{-1}}{b(E)\,\sqrt{\pi}}
\, \left[ \frac{r_s}{\lambda} \right]^{n\gamma}\;.
\een
As expected, the electron density scales like $\lambda^{-n\gamma}$, where 
$\lambda$ is the diffusion scale, but we have grounded this intuition on a 
rather rigorous demonstration. Of course, this approximation is not valid
for vanishingly small values of $\lambda$. This means that at any energy
but the maximal energy injected from DM annihilation or decay --- when 
$\lambda$ is sizable --- the actual electron density is not governed by the 
(squared) DM density itself, but is diluted away over a region delineated by 
$\lambda$.

This has important consequences for the SZ calculation. Indeed, we have
computed the optical depth in the diffusion-less approximation, showing 
that it was significantly increasing with angular resolution for cuspy
profiles. Nevertheless, we have just demonstrated that the electron density 
should in fact saturate within a scale fixed by $\sim \lambda$, leading to the 
saturation of the optical depth itself, independently of the angular 
resolution. For $\lambda\sim 10$ kpc, this would correspond to an 
angular size of $\sim 40''$ for our template cluster, as depicted in 
Fig.~\ref{fig:res}, which means that instruments with better resolution than 
this value would not improve the detection potential.

The full SZ computation is slightly different when spatial diffusion is
considered, since we can no longer factorize the energy and the spatial parts 
as we did in Eqs.~(\ref{eq:dnde_dm},\ref{eq:nel_noav0},\ref{eq:nel_noav1}), 
but must use Eq.~(\ref{eq:exact_ne}) to compute the the optical depth. 
Nevertheless, up to a good approximation valid in the monochromatic injection 
case, we can associate an energy-dependent core radius of size 
$\lambda(E,n\,m_\chi/2)$ to each halo model. Given the values that we used for 
the diffusion coefficient and for the energy loss rate, this would lead to a 
plateau in the optical depth curves for scales below $\sim 10$ kpc, as also 
found in the final published version of~\cite{2009JCAP...10..013Y}. Since our 
spectral analysis was performed for an NFW profile and an angular resolution 
of $10''$, the full spatial diffusion calculation would not only decrease the 
predicted amplitude by a factor of $\sim 2$, as seen from Fig.~\ref{fig:res}, 
but also make this result constant for any better angular resolution. Should 
have we done this exercise with a Moore profile, we would have overshot the 
actual amplitude by one order of magnitude.
%% JL

\section{Conclusion}
\label{sec:concl}

In this paper, we have revisited the prospect for radio experiments, 
like Planck or ALMA, for observing SZ distortions of the CMB spectrum due 
to the presence of relativistic electrons induced by DM annihilation or decay. 
Whenever possible, we have provided analytical grounds to the discussion.

In Sec.~\ref{sec:dm_el}, we have first focused on the most important physical 
weight on the distortion amplitude, the optical depth associated with the 
electron density (the thermal background as well as electrons induced by DM 
annihilation or decay). We have considered typical values for the DM particle 
properties and a generic template cluster located at a distance of 50 Mpc 
(Coma lies at $\sim$ 90 Mpc), with features very close to those observed in 
N-body simulations and a thermal electron population with a density of 
$n_{e,{\rm th}}=0.01\,{\rm cm^{-3}}$, consistent with most of X-ray observations.

In the frame of the diffusion-less limit proposed in \cite{2006A&A...455...21C},
we have stressed the importance of the angular resolution inside which the SZ 
signal has to be averaged, showing that even for very optimistic values of 
$\sim 1''$, the thermal electron optical depth almost always strongly dominate 
over that of the DM-induced electrons. This actually means that the thermal SZ 
completely overcomes the DM SZ at all frequencies but the thermal transition 
frequency. In addressing the angular 
resolution issue, we have derived simple and useful analytical approximations 
of the line-of-sight integral (Eqs.~\ref{eq:sph_approx} and~\ref{eq:los_approx},
the latter being more accurate) that can be used to compute the optical depth 
for any decaying or annihilating DM model. Actually, they are also useful for 
quick predictions of $\gamma$-ray fluxes from DM annihilation or decay in any
extragalactic objects, subhalos, dwarf spheroidals, galaxies or clusters.

Then, we have performed the full spectral analysis of the SZ distortion 
due to DM-induced electrons in Sec.~\ref{sec:res}, showing that it can 
hardly exceed the $\mu$K level in terms of temperature fluctuations, even 
for \textit{e.g.} very light DM particles, which are further constrained by 
511 keV observations of the Galactic center. Unconventional
very cuspy cluster halo profiles and very large annihilation cross sections or 
decay rates that would escape other astrophysical contraints for some reasons 
might still lead to signal close to future sensitivities. However, such 
configurations are not supported by current studies in each of the specific 
inputs and are therefore quite unlikely.

Finally, we have discussed the impact of considering the spatial transport
of electrons in clusters in Sec.~\ref{sec:diff}, providing again analytical
results. We have demonstrated that the diffusion-less limit is actually not 
valid in the cluster center for SZ computations with small angular resolutions 
because the electron density is actually smoothed over a scale set by the 
transport scale. This scale is small only for energies close to their injection 
values, and are therefore sizable over most the whole energy range but for 
$E\rightarrow \mchi $ in the case of annihilating DM. Given the propagation 
parameters used in our analysis, this scale is of order $\sim$ 
10 kpc, below which the optical depth saturates. This means that improving
the angular resolution below that scale will not improve the detection
potential, unfortunately. Our results are in good agreement with those 
obtained in~\cite{2009JCAP...10..013Y}, to which they provide deeper 
analytical grounds. Reversely, we cannot support the statements 
made in~\cite{2001ApJ...562...24C,2004A&A...422L..23C,2006A&A...455...21C} 
about the promising potential of SZ observations for indirect detection of DM.

\acknowledgments{
We are indebted to M. Langer and M. Fairbairn, who initiated the early stages 
of this work with us. JL is grateful to LAPTH for hospitality during different 
stages of this study, and acknowledges financial support by the 
French ANR project ToolsDMColl (BLAN07-2-194882).}

\bibliography{lavalle_bib}
\bibliographystyle{JHEP}

\end{document}